\documentclass[preprint,11pt]{imsart}

\usepackage{amsthm,amsmath,amsfonts,enumerate,graphicx,mathtools}
\RequirePackage[colorlinks,citecolor=blue,urlcolor=blue]{hyperref}

\startlocaldefs
	
	\newcommand{\PP}{\mbox{P}}
	\newcommand{\EE}{\mathbb{E}}
	\newcommand{\jasper}{\textcolor{black}}
	\newcommand{\juan}{\textcolor{black}}
	\newcommand{\jasperv}{\color{black}}
	\newcommand{\BB}{\textbf}
	\DeclareMathOperator*{\argmin}{arg\,min}
	
	\numberwithin{equation}{section}
	\theoremstyle{plain}
	\newtheorem{thm}{Theorem}[section]
	\theoremstyle{remark}
	\newtheorem{rem}{Remark}[section]
\endlocaldefs

\begin{document}

\begin{frontmatter}

\title{Improving precipitation forecasts using extreme quantile regression}

\runtitle{Extreme quantile regression}

\begin{aug}
\author{\fnms{Jasper} \snm{Velthoen}\thanksref{a,e1} \corref{} \ead[label=e1,mark]{j.j.velthoen@tudelft.nl}}
\author{\fnms{Juan-Juan} \snm{Cai}\thanksref{a,e2}\ead[label=e2,mark]{j.j.cai@tudelft.nl}}
\author{\fnms{Geurt} \snm{Jongbloed}\thanksref{a,e3}\ead[label=e3,mark]{g.jongbloed@tudelft.nl}}
\and
\author{\fnms{Maurice} \snm{Schmeits} \thanksref{b,e4}\ead[label=e4,mark]{maurice.schmeits@knmi.nl}}

\address[a]{Department of Applied Mathematics, Delft University of Technology, Mekelweg 4 2628 CD Delft \printead{e1,e2,e3}}
\address[b]{R\&D Weather and Climate Modelling, The Royal Netherlands Meteorological Institute (KNMI), Utrechtseweg 297 3731 GA De Bilt \printead{e4}}

\runauthor{J. Velthoen et al.}

\affiliation{Delft University of Technology}

\end{aug}

\begin{abstract}
Aiming to estimate extreme precipitation forecast quantiles, we propose a nonparametric regression model that features a constant extreme value index. Using local linear quantile regression and an extrapolation technique from extreme value theory, we develop an estimator for conditional quantiles corresponding to extreme high probability levels. We establish uniform consistency and asymptotic normality of the estimators. In a simulation  study, we examine the performance of our estimator on finite samples in comparison with a method assuming linear quantiles. On a precipitation data set in the Netherlands, these estimators have greater predictive skill compared to the upper member of ensemble forecasts provided by a numerical weather prediction model.
\end{abstract}


\end{frontmatter}

\section{Introduction}

{\jasperv
Extreme precipitation events can cause large economic losses, when large amounts of water cannot be properly drained. For example, water boards in the Netherlands, responsible for water management, need to take preventive action in the case of large amounts of precipitation to prevent flooding. Accurate predictions are therefore vital for taking preventive measures such as pumping the water out of the system. 

Weather forecasting relies on deterministic forecasts obtained by numerical weather prediction (NWP) models \cite{Kalnay2003}. These models are based on non-linear differential equations from physics describing the flow in the atmosphere. Starting from an initial condition of the atmosphere and using so-called physical parametrizations to account for unresolved physical processes, the NWP models are used to forecast precipitation, among other weather quantities. } 

The uncertainty in these types of forecasts is attributed to uncertainty in the initial condition and in the physical parametrizations in the model itself. An ensemble prediction system quantifies the uncertainty due to these two factors by applying small perturbations to the original quantities and running the NWP model multiple times subsequently. An ensemble forecast is to be viewed as a sample from the distribution of the predicted variable, where uncertainties in initial condition and model parametrizations are taken into account. Therefore, it is natural to consider the empirical distribution function of the ensemble forecast as an estimator of the distribution of the predicted variable, in this paper precipitation.

While the NWP ensemble prediction systems are rather skilful in forecasting precipitation for relatively short lead times, skill quickly decreases as lead time increases. Using upper ensemble members for forecasting extreme precipitation appears to be most challenging, due to the large spatial and temporal uncertainties of precipitation forecasts. Most methods that have been proposed to post-process forecasts are instead focussed on the bulk of the conditional distribution, see \cite{Wilks2011}. 

For the upper ensemble members there are two serious problems. First, the upper ensemble members tend to be not well calibrated, i.e. not reliable \cite{Bentzien2014}, especially for large amounts of precipitation, this is shown in \cite{Bentzien2012}. Second, the highest probability level of the extreme precipitation forecast is limited by the number of ensemble members, which is typically not large due to computational costs. In the ensemble prediction system of the European Centre for Medium-Range Weather Forecasts (ECMWF), which we consider in our case study, the system generates 51 ensemble members. Thus, the largest probability level is given by $\frac{51}{52}$.

In this paper, we aim to develop a post-processing approach for predicting extreme precipitation quantiles. More precisely, we focus on the problem of estimating the tail of the conditional distribution $F_{Y|X}$, with  $X$ a precipitation forecast by the NWP model and $Y$ the observed precipitation. We are interested in the function $x \mapsto Q_{Y|X}(\tau|x)$ for  $\tau$ close to one, where $Q_{Y|X}$ denotes the conditional quantile function.

Several estimators have already been proposed to estimate extreme conditional quantiles. All these estimators have a similar  structure consisting of two steps. First, the quantile function $Q_{Y|X}$ is estimated for moderately high probability levels $\tau$. In the second step, these estimated quantiles are used to extrapolate to obtain estimators of extreme conditional quantiles.

For the first step, general quantile estimation techniques are used. Examples are linear quantile regression in \cite{Wang2012} and \cite{Wang2013}, a local polynomial approximation to the quantile function \cite{Beirlant2004}, a $k$-nearest neightbour approach in \cite{Gardes2010} and inverse of empirical conditional distribution functions smoothed in the covariates in  \cite{Daouia2011}, and \cite{Daouia2013}.
For the second step two `types' of approaches can be distinguished. First, a local approach, where an extreme quantile estimator is applied to a sequence of estimated quantiles for moderately high probability levels attained from the first step. This method is used in \cite{Wang2012}, \cite{Wang2013}, \cite{Daouia2011}, \cite{Daouia2013}, \cite{Gardes2010}, \cite{Goegebeur2016} and \cite{Gardes2017}. The second type, where the exceedances above a threshold estimated in the first step are used to fit a generalized Pareto distribution, was introduced in \cite{Davison1990}. An application of the result of \cite{Davison1990} to precipitation data is discussed in \cite{Bentzien2012}, where a generalized Pareto distribution is fitted to the exceedances above an estimated linear quantile. They showed skilful short-range forecasts of extreme quantiles.

Most methods allow for a varying extreme value index depending on the covariates. The estimators of extreme value indices in such models are generally subject  to high variability. In the context of weather forecasting, this may lead to inconsistent forecasts over the covariates. 
After carefully considering the trade-off between the generality of the model and the efficiency of the estimation, we propose an additive model with a constant extreme value index for all covariates  cf. \eqref{eq:model}. {\juan Moreover, we assume that the extreme value index is positive. This assumption is supported by the result of our empirical study on summer rainfall in the Netherlands as well as the existing literature on precipitation data including \cite{ColesTawn1996}, \cite{Buishandetal2008} and \cite{GardesGirard2010}. Apart from this, our model assumes that the conditional quantile of $Y$ is a non-parametric function of the covariate, thus no parametric structure is required.  In our two step procedure, we first estimate a non-stationary threshold, namely the non-parametric quantile function by local linear quantile regression and then extrapolate to extreme quantiles based on the exceedances of this threshold.}


The scientific contribution of this paper is fourfold. First, we propose a model that achieves a good balance between generality and estimation efficiency and it fits the feature of post-processing data sets. Second, we derive asymptotic properties of the estimators, by first showing uniform consistency of local linear quantile regression, using a uniform Bahadur representation for the quantile estimator. Moreover, we establish asymptotic normality of the estimators of the extreme value index as well as the extreme conditional quantiles. Third, we address the issues such as selection of the bandwidth and tuning parameters, which is highly relevant from the application point of view. \jasper{Fourth, our procedure yields skilful prediction outperforming the upper ensemble member and showing similar skill to the linear estimator \cite{Wang2012} based on cross-validation}. Besides, our procedure can extrapolate to an extreme probability level that goes beyond the empirical quantile associated with the upper ensemble member. 

The outline of the paper is as follows: Section \ref{sec:model} we present our proposed model and develop the estimating procedures. The asymptotic properties of the estimator are studied in Section \ref{sec:asymp}. In Section \ref{sec:bandwidth} we propose a data driven approach for bandwidth selection. We show with a detailed simulation study in \jasper{Section \ref{sec:simulation} the finite sample performance of our estimator and compare it with an existing method. In Section \ref{sec:case} we apply our estimator to a dataset of precipitation observations and ensemble forecasts in the Netherlands. Finally, in Section \ref{sec:discussion} we discuss future research directions. The proofs of the theoretical results are provided in the appendix.}

\section{Model and Estimation} \label{sec:model}
We aim  to estimate the conditional tail quantiles of $Y$ given $X$, namely $Q_{Y|X}(\tau|\cdot)$ for $\tau$ close to one. To this end, we assume that there exists a $\tau_c \in (0,1)$ such that 
\begin{equation} \label{eq:model}
		Q_{Y|X}(\tau|x) = r(x) + Q_{\epsilon}(\tau) \mbox{  if } \tau \geq \tau_c,
\end{equation}
where  $r$ is a smooth continuous function and $Q_{\epsilon}$ denotes the quantile function of an error variable $\epsilon$, which is independent of $X$.  In order to make the model identifiable, it is assumed that $Q_{\epsilon}(\tau_c) = 0$. As a result, $Q_{Y|X}(\tau_c|x) = r(x)$. Moreover, we assume that the distribution of $\epsilon$ has a heavy right tail, that is there exists $\gamma > 0$ such that,
\begin{equation} \label{eq:doa}
	\lim_{t \to \infty} \frac{Q_{\epsilon}\left(1-\frac{1}{tx}\right)}{Q_{\epsilon}\left(1-\frac{1}{t}\right)} = x^{\gamma} \mbox{ ,  } x>0,
\end{equation}
where $\gamma$ is the extreme value index of $\epsilon$. {\juan {Note that \eqref{eq:doa} implies that the conditional distribution of $Y$ given that $X=x$ also has a heavy right tail with the same extreme value index $\gamma$.}}

 It is important to note that this additive structure is only assumed for probability levels $\tau$ exceeding $\tau_c$, which allows us to model the tail of the conditional distribution without assuming structure for $\tau<\tau_c$.
On one hand,  the quantile curve  $x \to Q_{Y|X}(\tau|x)$   for any $\tau \geq \tau_c$ has the same shape
 as $r$. On the other hand, the distance between the two quantile curves, that is 
$Q_{Y|X}(\tau_1|x) - Q_{Y|X}(\tau_2|x)$ for any $\tau_1>\tau_2\geq \tau_c$, is determined by $Q_{\epsilon}$ only and thus does not depend on $x$. We will refer to our model as   the Common Shape Tail (CST) model.
 
We remark that various types of additive structures have been proposed in recent studies on modeling extremes with covariates.  In \cite{Wang2012}, a  linear structure is assumed for $r$, where two scenarios are considered: the slope of the linear function is a nonparametric function of $\tau$ or it is constant. The latter scenario is a special case of our model. In \cite{Wang2013}, a linear structure is assumed for the conditional quantile function after the power transformation.  In both papers, $r$ is estimated by linear quantile regression. In \cite{Martins2016},  a nonparametric location-scale representation is assumed and local linear mean regression is used to estimate the conditional quantile called  $\alpha$-CVaR in that paper, where the existence of the fourth moment of the error variable is required. This requirement implies an upper bound on the extreme value index: $\gamma<\frac{1}{4}$.

Let $(X_1,Y_1), \ldots , (X_n,Y_n)$ denote i.i.d. paired observations satisfying \eqref{eq:model}. Based on this random sample, we construct a two step estimation procedure for ${Q}_{Y|X}(\tau_n|\cdot)$, where for asymptotics, $\tau_n\rightarrow 1$ as $n\rightarrow \infty$. We shall estimate $r$ and $Q_{\epsilon}(\tau_n)$ respectively in each of the two steps.

First, for the estimation of $r$ we choose to follow the local linear quantile regression approach studied in \cite{Yu1998}. An obvious advantage of the quantile regression approach is that it does not impose a constraint on the moments of the conditional distribution. 
 Let $h = h_n$ denote the bandwidth. In a window of size $2h$ around a fixed point $x$, we approximate the function linearly:  
$$
r(\tilde{x}) \approx r(x) + r'(x)(\tilde{x}-x)=:\alpha+\beta(\tilde x-x),~~~~~~\tilde x\in[x-h, x+h].
$$
The function $r$ and its derivative are estimated by the solution of the following minimization problem:
\begin{equation} \label{eq:estimator-r}
		(\hat{r}_n(x),\hat{r}_n'(x)) = \argmin_{(\alpha,\beta)} \sum_{i=1}^n \rho_{\tau_c}(Y_i-\alpha - \beta(X_i - x))K\left(\frac{X_i - x}{h}\right),
\end{equation}
	where $\rho_{\tau}(u) = u(\tau-I(u<0))$ is the quantile check function, cf.  \cite{Koenker2005} and $K$ a symmetric probability density function with $[-1,1]$ as support.
	
Second, for the estimation of $Q_{\epsilon}(\tau_n)$, we consider the residuals defined by $e_i = Y_i - \hat{r}_n(X_i)$, $ i =1,\ldots, n$.  Using the representation of  $Y_i = Q_{Y|X}(U_i|X_i)$, with $\{U_i, i=1,\ldots, n\}$ i.i.d. uniform random variables, and the model assumption \eqref{eq:model}, the residuals permit a more practical expression as below. 
\begin{equation}
	\label{eq:residuals}
		e_i = 
		\begin{cases}
			Q_{\epsilon}(U_i) + (r(X_i) - \hat{r}(X_i)) &\mbox{ if } U_i\geq \tau_c\\
			Q_{Y|X}(U_i|X_i) - \hat{r}(X_i) &\mbox{ otherwise }.		
		\end{cases}
\end{equation}

Denote the order statistics of the residuals by $e_{1,n} \leq \ldots \leq e_{n,n}$. Let  $k_n$ be an intermediate sequence depending on $n$ such that $k_n \to \infty$ and $k_n/n \to 0$ as $n \to \infty$. Then a Hill estimator of the extreme value index is given by
\begin{equation*}	
 \hat{\gamma}_n = \frac{1}{k_n}\sum_{i=1}^{k_n} \log \frac{e_{n-i+1,n}}{e_{n-k_n,n}}.
\end{equation*} 

The intuitive argument behind this estimator is that $\{e_{n-i,n}, i=0,\ldots, k_n\}$ are
 asymptotically equivalent to the upper order statistics of a random sample from the distribution of $\epsilon$, i.e. for some $\delta > 0$,
\begin{equation*} 
	\max_{i=0,\ldots,k_n} |e_{n-i,n} -  Q_{\epsilon}(U_{n-i,n})| = o_p(n^{-\delta});
\end{equation*}
see  the proof of Theorem \ref{thm:gamma} in the Appendix. For the same reason, we use the well known Weissman estimator of ${Q}_{\epsilon}(\tau_n)$ based on the upper residuals: 
\begin{equation} \label{eq:epsilon-estimator}
	\hat{Q}_{\epsilon}(\tau_n) = e_{n-k_n,n}\left(\frac{k_n}{n(1-\tau_n)}\right)^{\hat{\gamma}_n}.
\end{equation}

Combining the estimator of $r(x)$ given by \eqref{eq:estimator-r}  and the estimator of ${Q}_{\epsilon}(\tau_n)$ given by  \eqref{eq:epsilon-estimator}, we obtain the estimator of the conditional tail quantile:
\begin{equation} \label{eq:full-estimator}
	\hat{Q}_{Y|X}(\tau_n|x) = \hat{r}(x) + \hat{Q}_{\epsilon}(\tau_n).
\end{equation}
By construction, this estimator of the conditional tail quantile is continuous in $x$. We shall refer to our estimator as CST-estimator.

\section{Asymptotic Properties} \label{sec:asymp}

In this section, we present the asymptotic properties of the estimators obtained in Section \ref{sec:model}. 
We begin with uniform consistency of $\hat r_n$ in \eqref{eq:estimator-r}. 
We first state the assumptions with respect to our model \eqref{eq:model}. Let $g$ denote the density of $X$,  $f_{Y|X}(\cdot|x)$ denote the conditional density of $Y$ given $X=x$ and $c$ denote an arbitrary finite constant.
\begin{enumerate}[{A}1]
	\item The support of $g$ is given by $[a,b]$ and $\sup_{x \in [a,b]} |g'(x)| \leq c$.
	\item The third derivative of $r$ is bounded, i.e. $\sup_{x\in [a, b]}|r'''(x)|\leq c$.
	\item The function $x\to f_{Y|X}(r(x)|x)$ is Lipschitz continuous and $f_{Y|X}(r(x)|x) > 0$ for all $x\in[a,b]$.
\end{enumerate}

	
	\begin{thm}
	\label{thm:consistency}
		Let $\hat{r}_n$ be the estimator defined in \eqref{eq:estimator-r}. Choose $K$ a symmetric Lipschitz continuous probability density function supported on $[-1,1]$ and $h_n = O(n^{-\delta_h})$, with $\delta_h \in \left( \frac{1}{5},\frac{1}{2} \right)$. Under Assumptions $\mathrm{A1}$-$\mathrm{A3}$, there exists a $\delta \in (0, \frac{1}{2}-\delta_h)$ such that as $n\to \infty$,
			\begin{equation*}
				\sup_{x \in [a,b]} |\hat{r}_n(x) - r(x)| = o_p(n^{-\delta}). 
			\end{equation*}
	\end{thm} 
This theorem quantifies  the direct estimation error made in the first step of our procedure. Note that  the ``error" made in the first step  is transmitted to the second step by the definition of the residuals. Thus, the uniform consistency of $\hat r$ is important for deriving  the asymptotic property of $\hat{Q}_{Y|X}(\tau_n|\cdot)$ not only because $\hat r$ is a constructing part of  $\hat{Q}_{Y|X}(\tau_n|\cdot)$, but it also influences the asymptotic behavior of  $\hat{Q}_{\epsilon}(\tau_n)$.

\begin{rem}
Although many studies have been devoted to the non-parametric quantile regression, to the best of our knowledge, there is no existing result on the uniform consistency for $\hat r_n$ for an additive model.
In \cite{Kong2010}, a general uniform Bahadur  representation is obtained for local polynomial estimators of M-regression for a multivariate additive model. A local linear quantile regression is one of the M-regression and thus is included in the estimators considered in that paper. Corollary 1 in \cite{Kong2010} is our starting point for deriving the uniform consistency of $\hat r_n$.
\end{rem}	
	
For the asymptotic normality of $\hat{\gamma}_n$, we assume that $Q_{\epsilon}$ satisfies the following  condition, which is a second order strengthening of \eqref{eq:doa}. 
\begin{enumerate}[{A}4]
 \item There exist $\gamma > 0$, $\varrho < 0$ and an eventually positive or negative  function $A(t)$ with $\lim_{t\to \infty}A(t)=0$ such that for all $x>0$,
\begin{equation} \label{eq:sec-ord-cond}
	\lim_{t \to \infty} \frac{\frac{Q_{\epsilon}\left(1-\frac{1}{xt}\right)}{Q_{\epsilon}\left(1-\frac{1}{t}\right)}-x^{\gamma}}{A(t)} = x^{\gamma}\frac{x^{\varrho}-1}{\varrho}.
\end{equation}
\end{enumerate}
As a consequence, $|A(t)|$ is regularly varying with index $\varrho$. 

	\begin{thm} \label{thm:gamma}
		Let the conditions of Theorem \ref{thm:consistency} and $\mathrm{A4}$ be satisfied. Let $k_n \to \infty$ and $k_n/n \to 0$, {\juan{ $\sqrt{k_n}A(n/k_n) \to \lambda \in \mathbf{R}$ }}and $k_n^{\gamma + 1}n^{-(\delta + \gamma)} \to 0$ as $n\to\infty$, with $\delta$ from Theorem \ref{thm:consistency}. Then
		\begin{align*}
			\sqrt{k_n}(\hat{\gamma}_n - \gamma) \xrightarrow{d} N\left( \frac{\lambda}{1-\varrho},\gamma^2\right) \mbox{ as } n\to \infty.
		\end{align*}
	\end{thm}
\begin{rem}
When deriving asymptotic properties for extreme statistics, it typically requires some regular conditions on $k_n$, the number of tail observations used in the estimation when the sample size is $n$. For the original Hill estimator, which is based on i.i.d. observations, the asymptotic normality is proved under Assumption $\mathrm{A4}$ and $\sqrt{k_n}A(1-k_n/n) \to \lambda \in \mathbf{R}$.
 {\juan The  condition $\lim_{n\to \infty}k_n^{\gamma + 1}n^{-(\delta + \gamma)} = 0$  is used to make sure that  the upper order residuals behave similarly to the upper order statistics of a random sample from the distribution of $\epsilon$. Suppose one chooses $k_n=n^\alpha$ for $0<\alpha < \min\left(\frac{2\varrho}{2\varrho-1}, \frac{\delta+\gamma}{\gamma+1}\right)$,} it satisfies all the conditions on $k_n$. So in theory, there exists a wide range of choices for a proper $k_n$.\jasper{In practice, it is challenging to choose a $k_n$. In Section \ref{sec:simulation} we propose to use a fixed choice of $k_n$ that worked well in several simulation studies.}
\end{rem}

	The asymptotic normality of $\hat{Q}_{Y|X}(\tau_n|x)$  defined in \eqref{eq:full-estimator} is now given below. To simplify notation, we denote with $p_n = 1-\tau_n$. 
	\begin{thm}\label{thm:quantile}
	 Let the conditions of Theorem \ref{thm:gamma} be satisfied. Assume $np_n = o(k_n)$, $|\log(np_n)| = o(\sqrt{k_n})$ and $\frac{\sqrt{k_n p_n^{\gamma}}}{n^{\delta}\log \left(\frac{k_n}{np_n}\right)} \to 0$, then as $n\rightarrow\infty$,
		\begin{align*}
			\frac{\sqrt{k_n}}{\log \left(\frac{k_n}{np_n}\right)Q_{\epsilon}(\tau_n)}\left(\hat{Q}_{Y|X}(\tau_n|x) - Q_{Y|X}(\tau_n|x)\right) \xrightarrow{d} N\left(\frac{\lambda}{1-\varrho},\gamma^2\right).
		\end{align*}
	\end{thm}

\begin{rem}
The condition $np_n = o(k_n)$ guarantees that the conditional quantile is an extreme one. It gives the upper bound for $p_n$. And the condition $|\log(np_n)| = o(\sqrt{k_n})$ gives the lower  bound on $p_n$, which limits the range of extrapolation. Clearly $p_n=O(n^{-1})$ satisfies both conditions. The asymptotic normality holds even for some $p_n<\frac{1}{n}$, which means it is beyond the range of the available data. In the weather forecast context, predicting the amount of precipitation so extreme that it never occurred during  the observed period is also feasible. The assumption $\lim_{n\to \infty}\frac{\sqrt{k_n p_n^{\gamma}}}{n^{\delta}\log \left(\frac{k_n}{np_n}\right)} = 0$ is a technical condition we use to guarantee that the error made in the first step does not contribute to the limit distribution. 
\end{rem}

The proofs for Theorems \ref{thm:consistency}, \ref{thm:gamma} and \ref{thm:quantile} are provided in the Appendix.

\section{Bandwidth selection} \label{sec:bandwidth}

The selection of the bandwidth is a crucial step in local linear quantile regression cf. \eqref{eq:estimator-r}. The bandwidth controls the trade-off between the bias and variance of the estimator. Increasing the bandwidth $h$ decreases the variance, but tends to increase the bias due to larger approximation errors in the local linear expansion.

In \cite{Yu1998}, the authors propose to estimate the optimal bandwidth for quantile regression by rescaling the optimal bandwidth for mean regression. There is a rich literature on bandwidth selection for mean regression. However,   in our setting this approach is not satisfactory because the scaling factor is difficult to estimate and it also assumes the existence of the first moment, i.e. it limits us to the case $\gamma < 1$. 


Instead we adopt a bootstrap approach, similar to the one proposed in \cite{Beirlant2004} to estimate the  global optimal bandwidth with respect to the mean integrated squared error (MISE), i.e., 
\begin{equation*} 
	h_{opt} = \argmin_h \EE \left[ \int_a^b \left(Q_{Y|X}(\tau_c|x) - \hat{Q}_{Y|X}^h(\tau_c|x)\right)^2 \mbox{d}x\right]  =: \argmin_h S(h),
\end{equation*}
where $\hat{Q}_{Y|X}^h(\tau_c|x)$ denotes the $\tau_c$ quantile estimated by \eqref{eq:estimator-r} with bandwidth $h$.

Let $B$ denote the number of bootstrap samples. The bootstrap samples\\ $(X_1^j,Y_1^j), \ldots , (X_n^j,Y_n^j)$ for $j = 1,\ldots , B$ are sampled with replacement from the original $n$ data pairs. The optimal bandwidth is estimated by minimizing the bootstrap estimator $\hat{S}(h)$ of $S(h)$, which is given by the objective function in \eqref{eq:MISE-boot}.
\begin{equation} \label{eq:MISE-boot}
	\hat{h} = \argmin_h \frac{1}{B}\sum_{j=1}^B \int_a^b \left(\hat{Q}_{Y|X}^{h_0}(\tau_c|x) - \hat{Q}_{Y|X}^{h,j}(\tau_c|x)\right)^2 \mbox{d}x,
\end{equation} 
where $h_0$ is an initial bandwidth chosen by visual inspection and $\hat{Q}_{Y|X}^{h,j}(\tau_c|x)$ denotes the estimate of the conditional quantile function based on the $j$-th bootstrap sample.  In practice, the integral is approximated using numerical integration. 

Two alternative approaches were attempted. First, a bootstrap approach, fixing the covariates $X$ and sampling for each covariate level an uniform random variable $U$. For values of $U \geq \tau_c$ a positive residual $e$ is sampled and the bootstrap sample is $Y^b = \hat{Q}_{Y|X}^{h_0}(X) + e$. In the case $U<\tau_c$ a local linear quantile estimate is obtained at the covariate level $X$ with bandwidth $h_0$ at probability level $U$. The bandwidth is then estimated by the solution of the minimization in \eqref{eq:MISE-boot}. Second, a leave-one-out cross validation approach that minimizes the quantile loss function is used to obtain the estimator of the optimal bandwidth: 
\begin{equation*}
	\hat{h} = \argmin_h \hat{S}(h) = \argmin_h \sum_{i=1}^n \rho_{\tau_c}(Y_i - \hat{Q}_{Y|X}^{h,-i}(\tau_c|X_i)),
\end{equation*}
where  $\hat{Q}_{Y|X}^{h, -i}$ denotes the conditional quantile estimate with bandwidth $h$ and leaving out the $i$th observation.
Intuitively, the cross validation approach is attractive as it is much faster compared to the bootstrap approach and it is based on the idea of scoring the quantile curve with the same scoring function used for estimation. Yet, based on a simulation study, the direct bootstrap procedure performed significantly better compared to these alternative approaches. This is in accordance with the conclusions drawn in \cite{Beirlant2004}.

\section{Simulation} \label{sec:simulation}
In this section, the finite sample performance of the CST-estimator is assessed using a detailed simulation study. \jasper{A comparison is made with the estimator proposed in \cite{Wang2012}, where also a two step procedure is used. The first step consists of estimating a sequence of linear quantile curves for moderately high probability levels, using quantile regression. And the second step then uses a Hill estimator for the extreme value index based on the estimated quantiles.} Extrapolation to the extreme quantiles is done by  a Weissman type estimator, similar to the one in \eqref{eq:epsilon-estimator}.

Define the simulation model from which the data is drawn by,
\begin{equation} \label{eq:simulation-model}
	Y = r(X) + \sigma(X)\epsilon.
\end{equation}
We choose $X$ uniformly distributed in $[-1,1]$ and independently, $\epsilon$ follows from a generalized Pareto distribution with $\gamma=0.25$, or a Student $t_1$ distribution.  For the function $\sigma$, we consider two cases: $\sigma(x) = 1$ and $\sigma(x)=\frac{4+x}{4}$. Note that for $\sigma(x)=1$, our model assumption \eqref{eq:model} is satisfied with $\tau_c = 0$. For $\sigma(x) = \frac{4+x}{4}$, our model assumption is not satisfied since the distribution of the additive noise depends on $x$, which allows us to study the robustness of the model assumptions.

We consider three choices for the function $r$: linear, nonlinear monotone and a more wiggly function,
\begin{equation*}
 r_1(x) =x \mbox{, } r_2(x)=\exp(x) \mbox{, } r_3(x) = \sin(2\pi x)(1-\exp(x))
\end{equation*} 
Performance is compared for two sample sizes : $n=500$ and $n=2500$.

The estimation of the quantile curves $x \mapsto Q_{Y|X}(\tau|x)$ with $\tau = 0.99$ and $\tau =0.995$ is assessed with an empirical estimator of the mean integrated squared error:
$\frac{1}{m}\sum_{i=1}^m\int_{-1}^1  (\hat{Q}^{(i)}_{Y|X}(\tau|x)-Q_{Y|X}(\tau|x))^2 dx$, where $m=500$ and $\hat{Q}^{(i)}_{Y|X}(\tau|x)$ denotes the estimate based on the $i$-th sample. \jasper{The integral is approximated by numerical integration.} Tables \ref{tab:GPD} and \ref{tab:student} report the estimated MISE for different models and different methods.

For the CST estimator, we choose $\tau_c = 0.5$ while the model holds for any $\tau_c\geq 0$. \jasper{Simulations show that the results are not sensitive to the level of $\tau_c$ that is chosen.} The value of $k$ is typically chosen by inspection at the point where the Hill plot, i.e $(k,\hat{\gamma}(k))$, becomes stable. \jasper{In the simulation study it is not possible to choose the stable point for every simulation. Therefore, we choose a fixed $k=[4n^{1/4}]$, where $[.]$ denotes the integer part. From simulations we see that the estimate becomes stable around this value of $k$.}



For the estimator in \cite{Wang2012}, it is proposed to choose \jasper{$k=[4.5n^{1/3}]$}. \jasper{Additionally, the probability sequence for which the linear quantile curves are estimated is given by, $\frac{n-k}{n}, \cdots , \frac{n-3}{n}$, trimming of the most extreme quantiles, $\frac{n-2}{n}, \cdots , \frac{n}{n}$. This is needed in order to obtain a Bahadur expression for the regression quantiles. In \cite{Wang2012} it is suggested to trim off $[n^{\eta}]$ observations, with $\eta \in (0,0.2)$. In our simulation trimming off the three most extreme probabilities gave the best performance.} The estimator allows for  varying extreme value indices as well as a constant extreme value index. A constant extreme value index is used as this is assumed in our setting. We refer to this estimator as the linear estimator.  The model assumption for this method is satisfied only when $r=r_1$, the linear case. 


For generalized Pareto errors, the mean integrated squared errors are shown in Table \ref{tab:GPD}. For the case $\sigma(x) = 1$, the CST estimator performs best, as expected, since the data follow the model assumption (\ref{eq:model}). For the case $\sigma(x) = \frac{4+x}{4}$ a similar conclusion can be drawn for $n=500$. Though, for a sample size of $2500$ the linear  estimator does slightly better. \jasper{The deviation from the model assumption clearly affects the behaviour of the CST estimator, but not the linear estimator. The difference between the methods becomes visible for larger sample sizes as the bias for the CST estimator starts to play a bigger role in the MISE.} 

For Student $t_1$ errors, the results are shown in Table \ref{tab:student}. For sample size $n=500$, the CST estimator has  smaller MISE for $\tau=0.99$ and larger MISE for $\tau=0.995$, in comparison with the linear estimator. For a larger sample size $n=2500$, the CST estimator outperforms the linear method. \jasper{For small sample size the $r$ is subject to high variance locally, this leads to errors in the residuals and as a result in the extreme value index. This is shown in the extrapolation to the $0.995$ quantile. When the sample size is larger this is not an issue, which leads to better performance of the CST estimator. The relative effect of the deviation from the model by choosing $\sigma(x) = \frac{4+x}{x}$ is lower now for a large $\gamma =1$. As a result the CST estimator performs better sometimes for large sample size and $\sigma(x) = \frac{4+x}{x}$. }


\jasper{
\begin{rem}
	The estimator that is proposed in \cite{Daouia2011} was also compared to the CST estimator and the linear estimator and was outperformed clearly in all instances by these methods, although it is the only method for which the model assumptions are satisfied for all settings. The procedure does not assume any structure in the data and it allows for  varying extreme value indices, which requires to estimate the extreme value index locally by using  a very limited amount of observations. As a consequence, the function $\hat \gamma(x)$  fluctuates heavily and it further creates large inaccuracies in the quantile  extrapolation. 
From the simulation result, it is clear that this method suffers severely from lack of efficiency for the sample sizes considered here. Therefore, the results were left out to focus on the comparison between the CST and the linear method.
\end{rem}
}

\begin{table}
\caption{\label{tab:GPD} Mean integrated squared errors based on samples from (\ref{eq:simulation-model}), with errors GPD($\gamma = 0.25$).}
\centering
\begin{tabular}{lllllll}
\hline \hline
 & & \multicolumn{2}{c}{$\sigma(x) = 1$} && \multicolumn{2}{c}{$\sigma(x) = \frac{4+x}{4}$}\\
 $r$ & method & 0.99 & 0.995 && 0.99 & 0.995  \\ 
  \hline && \multicolumn{5}{c}{$n=500$}\\ \hline
 	$r_1$ & CST & \BB{2.62} & \BB{9.16} &  & \BB{5.28} & \BB{14.42} \\ 
  	$r_1$ & linear & 9.04 & 18.53 &  & 7.75 & 15.66 \\
   \hline
	$r_2$ & CST & \BB{2.78} & \BB{9.51} &  & \BB{5.64} & \BB{15.04} \\ 
  	$r_2$ & linear & 8.69 & 18.92 &  & 8.57 & 18.47 \\ 
   \hline
	$r_3$ & CST & \BB{2.66} & \BB{8.01} &  & \BB{5.27} & \BB{13.95} \\ 
  	$r_3$ & linear & 9.05 & 18.83 &  & 8.98 & 18.55 \\
   \hline && \multicolumn{5}{c}{$n=2500$}\\ \hline
	$r_1$ & CST & \BB{0.64} & \BB{1.59} &  & 3.23 & 5.91 \\ 
 	$r_1$ & linear & 2.04 & 6.14 &  & \BB{1.88} & \BB{5.53} \\ 
   \hline
	$r_2$ & CST & \BB{0.71} & \BB{1.70} &  & 3.24 & 5.85 \\ 
  	$r_2$ & linear & 1.95 & 6.09 &  & \BB{1.86} & \BB{5.64} \\ 
   \hline
	$r_3$ & CST & \BB{0.75} & \BB{1.56} &  & 3.42 & 6.04 \\ 
  	$r_3$ & linear & 2.15 & 5.98 &  & \BB{2.12} & \BB{5.91} \\ 
   \hline
\hline
\end{tabular}
\end{table}

\begin{table}
\caption{\label{tab:student} Mean integrated squared errors $\times 10^{-2}$ based on samples from (\ref{eq:simulation-model}), with errors from Student $t_1$.}
\centering
\begin{tabular}{lllllll}
\hline \hline
& & \multicolumn{2}{c}{$\sigma(x) = 1$} && \multicolumn{2}{c}{$\sigma(x) = \frac{4+x}{4}$}\\
 $r$ & method & 0.99 & 0.995 && 0.99 & 0.995 \\ 
  \hline && \multicolumn{5}{c}{$n=500$}\\ \hline
	$r_1$ & CST & \BB{3.41} & 31.69 &  & \BB{3.33} & 28.83 \\ 
 	$r_1$ & linear & 4.69 & \BB{26.66} &  & 4.74 & \BB{26.82} \\ 
   \hline
	$r_2$ & CST & \BB{3.83} & 38.35 &  & \BB{4.40} & 43.25 \\ 
  	$r_2$ & linear & 5.19 & \BB{30.44} &  & 5.01 & \BB{29.81} \\ 
   \hline
	$r_3$ & CST & \BB{3.97} & 40.56 &  & \BB{3.44} & \BB{29.47} \\ 
  	$r_3$ & linear & 4.78 & \BB{27.62} &  & 5.32 & 30.30 \\  
   \hline && \multicolumn{5}{c}{$n=2500$}\\ \hline
	$r_1$ & CST & \BB{0.69} & \BB{5.14} &  & \BB{1.20} & \BB{6.49} \\ 
  	$r_1$ & linear & 1.30 & 10.68 &  & 1.35 & 10.94 \\ 
   \hline
	$r_2$ & CST & \BB{0.82} & \BB{5.98} &  & 1.26 & \BB{7.31} \\ 
  	$r_2$ & linear & 1.27 & 10.70 &  & \BB{1.24} & 10.31 \\ 
   \hline
	$r_3$ & CST & \BB{0.83} & \BB{6.03} &  & \BB{1.17} & \BB{7.10} \\ 
  	$r_3$ & linear & 1.38 & 11.30 &  & 1.32 & 10.90 \\
   \hline
\hline
\end{tabular}
\end{table}

\section{Post-processing extreme precipitation} \label{sec:case}

Our dataset consists of observations and ECMWF ensemble forecasts of daily accumulated precipitation at eight meteorological stations spread across the Netherlands (de Bilt, De Kooy, Twente, Eelde, Leeuwarden, Beek, Schiphol and Vlissingen). The data in this study is for the warm half year, namely 15th of April until 15th of October, in the years $2011$ till $2017$. The lead time is defined as the time between initialization of the ensemble run and the end of the day at $00$ UTC for which the forecast is valid. We consider lead times from 24 hours up till 240 hours with 12 hour increments. For each lead time and location the number of observations is about 1287.

For fixed lead time and location, an ensemble forecast consists of 51 exchangeable members, which can be seen as a sample from the distribution of precipitation, where the uncertainty in the initial condition and model parametrizations are accounted for. As a result, quantile estimates for probability levels $\frac{i}{52}$, for $1 \leq i \leq 51$, are given by the order statistics of the ensemble forecast. \jasper{Note that the precipitation observations are not used by the ensemble forecast as standard the amount of precipitation is set to zero at initialization of the NWP model.}

In practice, it is known that the upper ensemble member is not well calibrated in the sense that it leads to underestimation of the extremes, see \cite{Bentzien2012}. This is partly caused by a representatively error, because the forecast is a grid-cell average and the observation is a station point value. Statistical post-processing can correct this and other systematic errors \cite{Wilks2011}. \jasper{For long lead times, a forecast, especially the upper ensemble member loses all predictive skill, \cite{Bentzien2012}.} We show that, by applying the CST estimator, we can calibrate the upper ensemble member and obtain more skilful forecasts for short and long lead times. To relate to the notation of Section \ref{sec:model}, we denote the daily accumulated precipitation by $Y$ and the upper ensemble member by $X$.

\jasper{For each lead time we pool data from all eight locations. These locations are spread over the Netherlands and as most extreme events are caused by local deep convective showers, the observations can be considered approximately independent. We compare the performance of the ensemble method with the CST estimator as in \eqref{eq:full-estimator} and the linear estimator as explained in Section \ref{sec:simulation}.}

{\jasperv
As precipitation is often modelled using a point mass on $0$ for the dry days, we model the point mass using a logistic regression with as covariate the number of ensemble members equal to zero. The distribution function is then given by:

\begin{equation} \label{eq:data-dist}
	F_{Y|X}(y|x) = p_0(x) + (1-p_0(x))F_{Y|X,Y>0}(y|x)
\end{equation}

Where the quantiles are given by:

\begin{equation} \label{eq:data-quant}
	Q_{Y|X}(\tau|x) = 
	\begin{cases}
		0 &\mbox{ if } \tau \leq p_0(x)\\
		Q_{Y|X, Y > 0}\left(\frac{\tau -p_0(x)}{1-p_0(x)}\right) &\mbox{ if } \tau > p_0(x)
	\end{cases}
\end{equation}

We then apply the CST estimator to estimate $Q_{Y|X, Y > 0}$, where we choose $\tau_c = 0.95$. This choice is based on best validation score, as explained below, based on one year of data. The bandwidth $h$ is determined using the bandwidth selection method described in Section \ref{sec:bandwidth} and $k = [4n^{1/4}]$, the same as in the simulation study. Alternative to choosing $X$ as the upper ensemble member we have also considered other ensemble members and trimmed means of the ensemble members. Among these choices the upper ensemble member showed best performance.

For the linear method we do not incorporate the point mass as the method already takes this into account as all quantiles are estimated globally instead of the CST estimator, which estimates the quantiles in a local manner. Incorporating the point mass led to severely worse results for the linear method. The same hyper parameters were chosen as in Section \ref{sec:simulation}; changing these did not influence the results.

Note that for days that have a large point mass on 0 and the rescaled probability is not extreme, in these cases we just use a local linear quantile estimator as described in Equation \ref{eq:estimator-r} as the estimator of $Q_{Y|X, Y > 0}\left(\frac{\tau -p_0(x)}{1-p_0(x)}\right)$.
}

The predictive performance of a quantile estimator $\hat{Q}_i(\tau)$ can be quantified by the quantile verification score and visualized by the quantile reliability diagram, which are discussed in detail in \cite{Bentzien2014}. The quantile verification score is defined as $\mbox{QVS}_{\tau}(\hat{Q}) = \sum_{i=1}^n \rho_{\tau}(Y_i-\hat{Q}_i(\tau))$, where $\rho_{\tau}$ is the quantile check function. The score is always positive, where low scores represent good performance and high scores bad performance. In \cite{Bentzien2014} it is shown that the score can be decomposed in three components: uncertainty, reliability and resolution, where only the last two depend on the estimator itself. A reliable or calibrated forecast has the same distribution as the underlying distribution that is estimated.

The quantile reliability diagram visualizes the reliability of the forecast quantile by creating equally sized bins with respect to the forecast quantile and then graphing the empirical quantile of the corresponding observations in the bin against the mean forecast quantile in the bin. For the forecast to be reliable these points should lie on the line $y=x$.

It is natural to compare the predictive performance of a quantile estimator to some reference quantile estimator $\hat{Q}_{\mbox{ref}}$. For this we take the climatological empirical quantiles as the reference method, i.e. the empirical quantiles of the sample $Y_i$, $1\leq i \leq n$. Note that this is the simplest estimate we can obtain without making use of a numerical weather prediction model. The quantile verification skill score, given by $\mbox{QVSS}_{\tau}(\hat{Q}) = 1-\frac{\mbox{QVS}_{\tau}(\hat{Q})}{\mbox{QVS}_{\tau}\left(\hat{Q}_{\mbox{ref}}\right)}$, is a relative measure of performance compared to the reference method, taking values in $(0,1]$ when $\hat{Q}$ improves on $\hat{Q}_{\mbox{ref}}$ and values below zero when the opposite is true.

\begin{figure*}
	\includegraphics[width=0.45\textwidth]{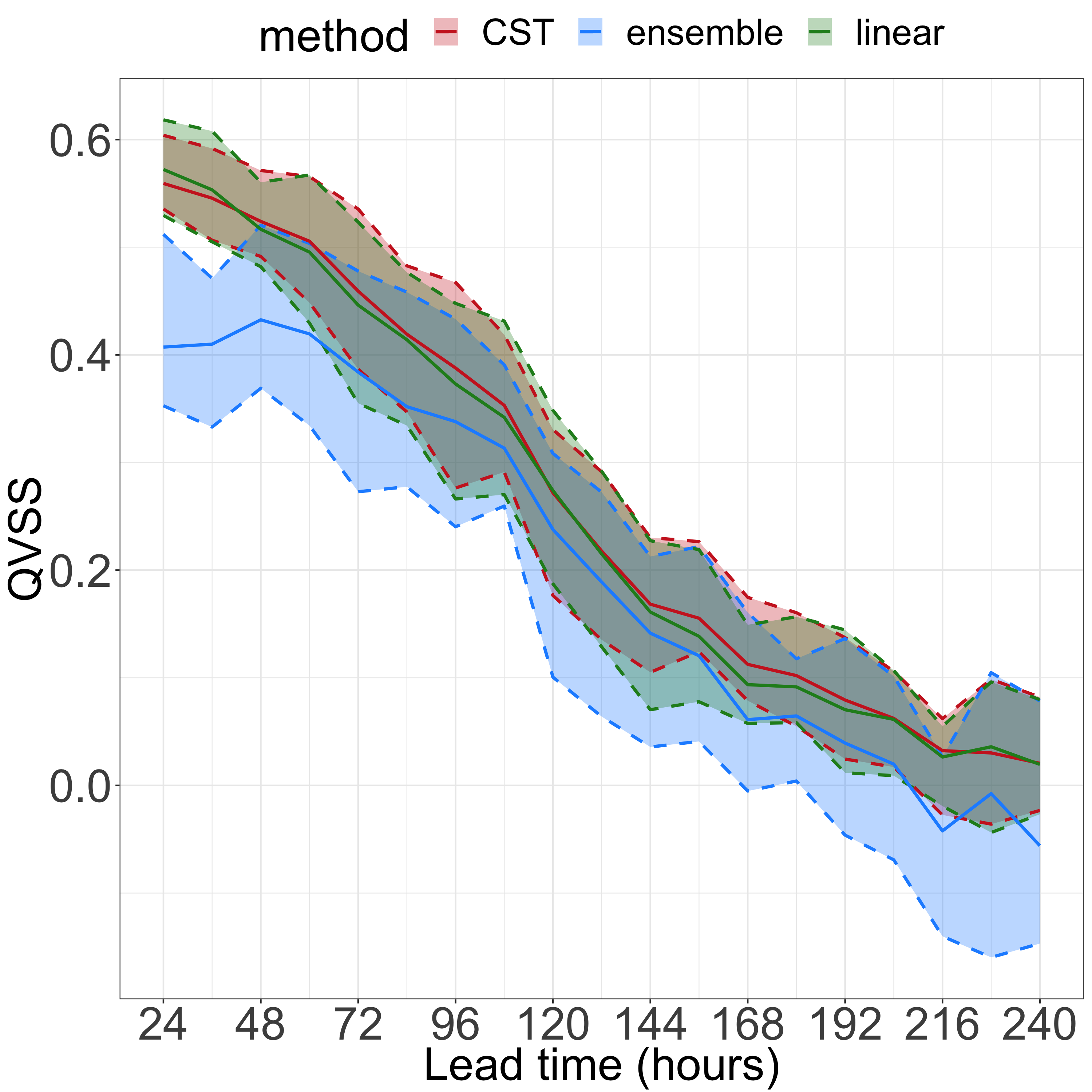}
	\includegraphics[width=0.45\textwidth]{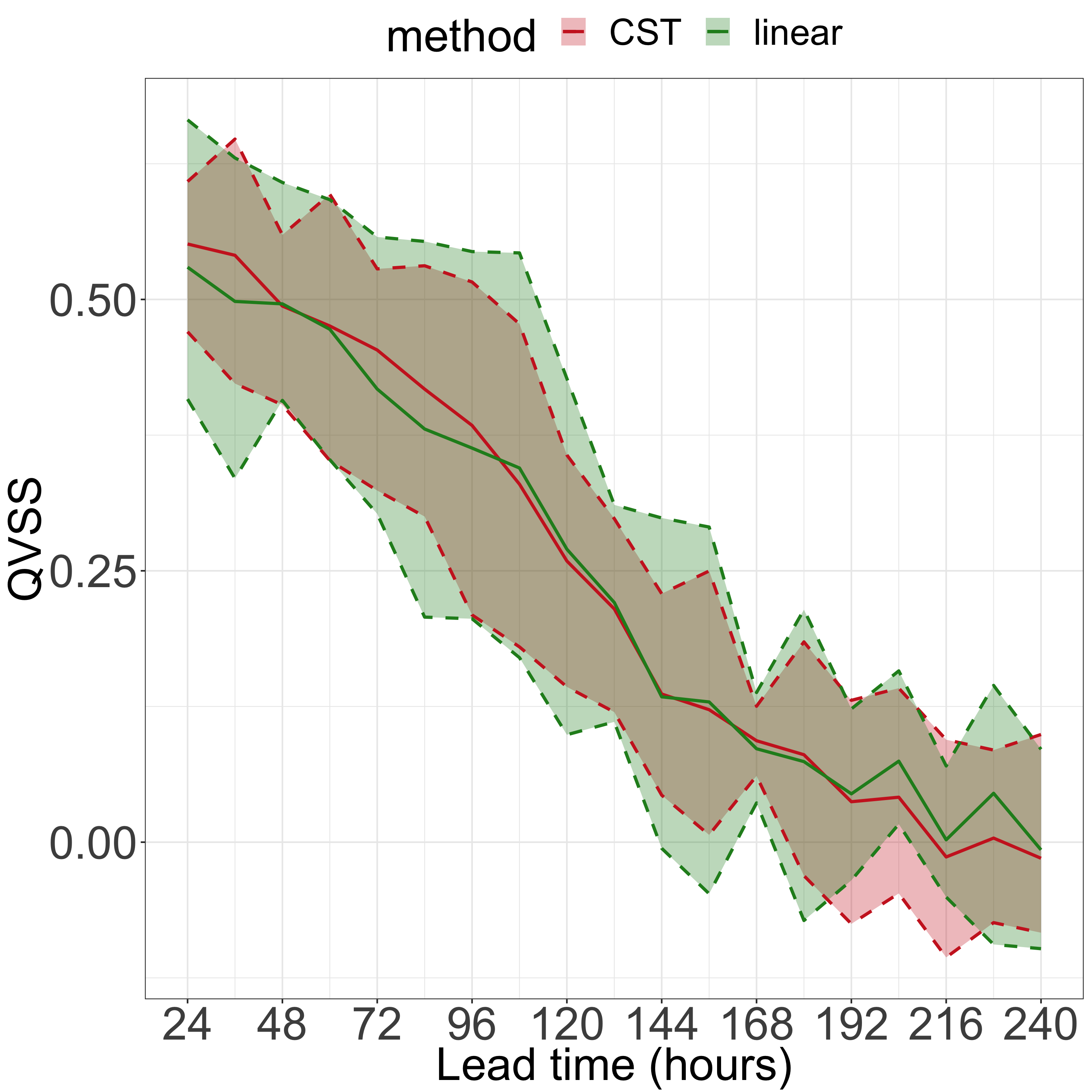}
	\caption{QVSS as a function of lead time for CST estimator in red, the ensemble in blue and the linear estimator in green, on the left for the $\frac{51}{52}$ quantile and on the right for the $0.995$ quantile. The bands are obtained by validating for each location separately.}
	\label{fig:data-QVSS}
\end{figure*}

\begin{figure*}
	\includegraphics[width=0.45\textwidth]{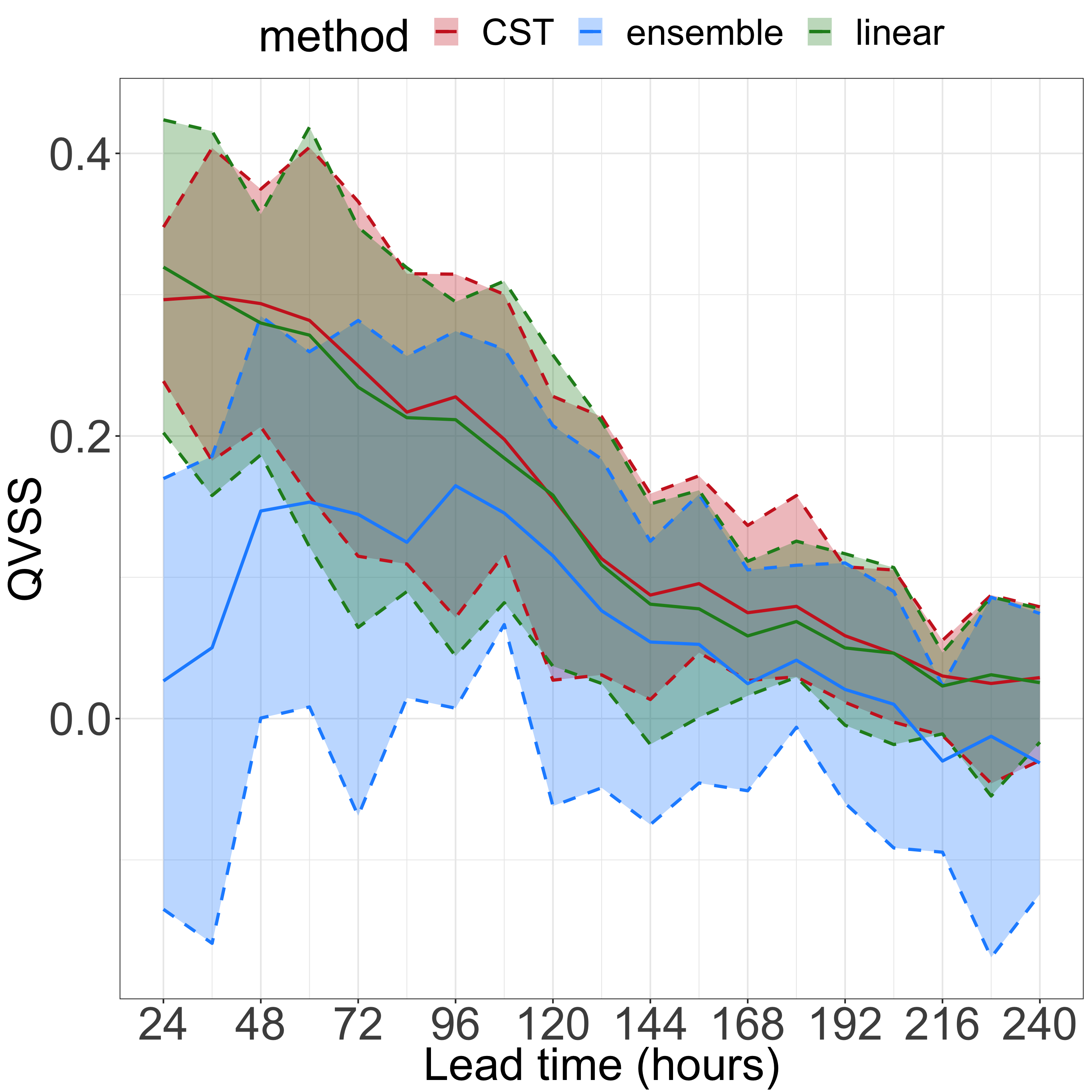}
	\includegraphics[width=0.45\textwidth]{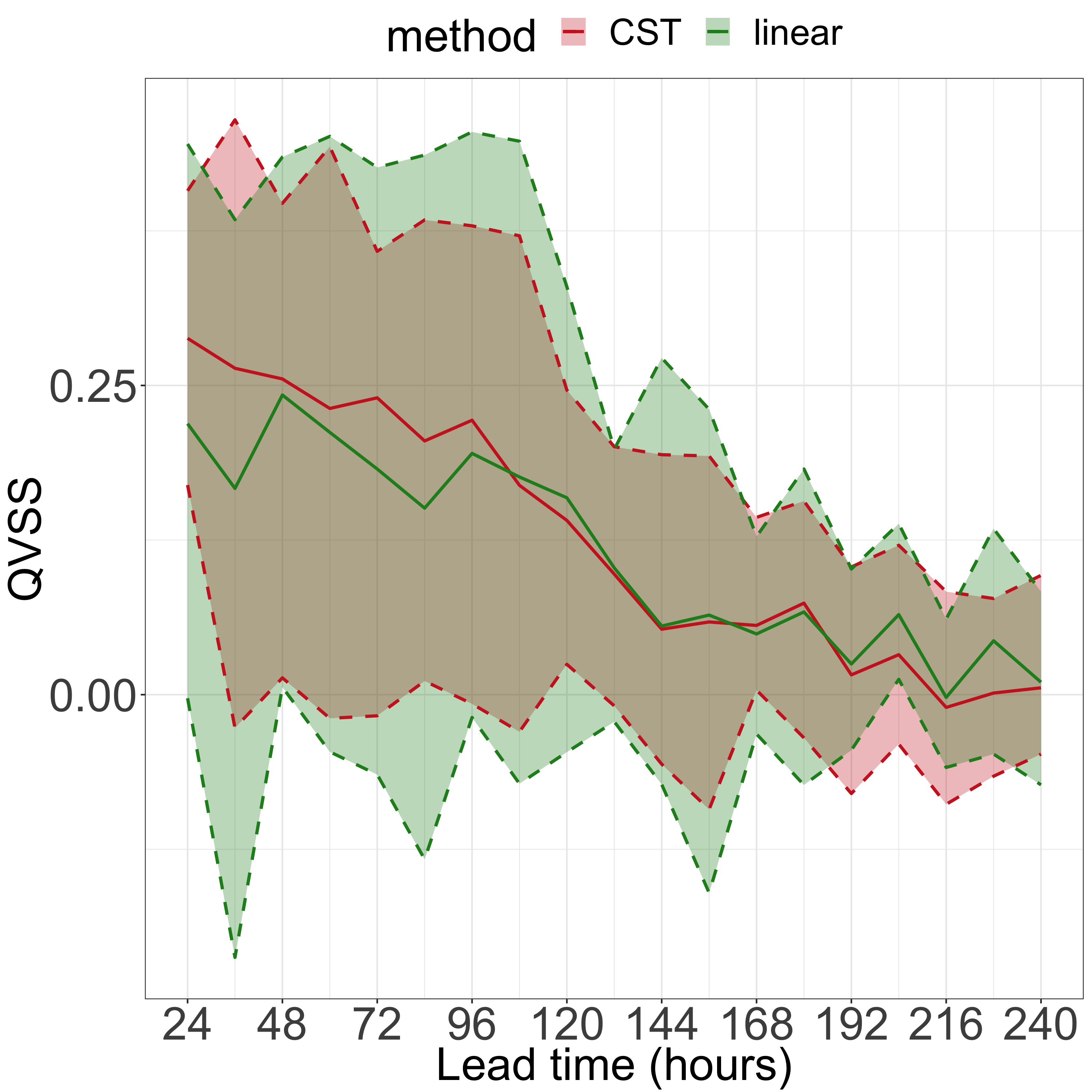}
	\caption{QVSS as a function of lead time conditioned on $X>5$, for CST estimator in red, the ensemble in blue and the linear estimator in green, on the left for the $\frac{51}{52}$ quantile and on the right for the $0.995$ quantile. The bands are obtained by validating for each location separately.}
	\label{fig:data-QVSS-thresh}
\end{figure*}

\begin{figure*}
	\includegraphics[width=0.45\textwidth]{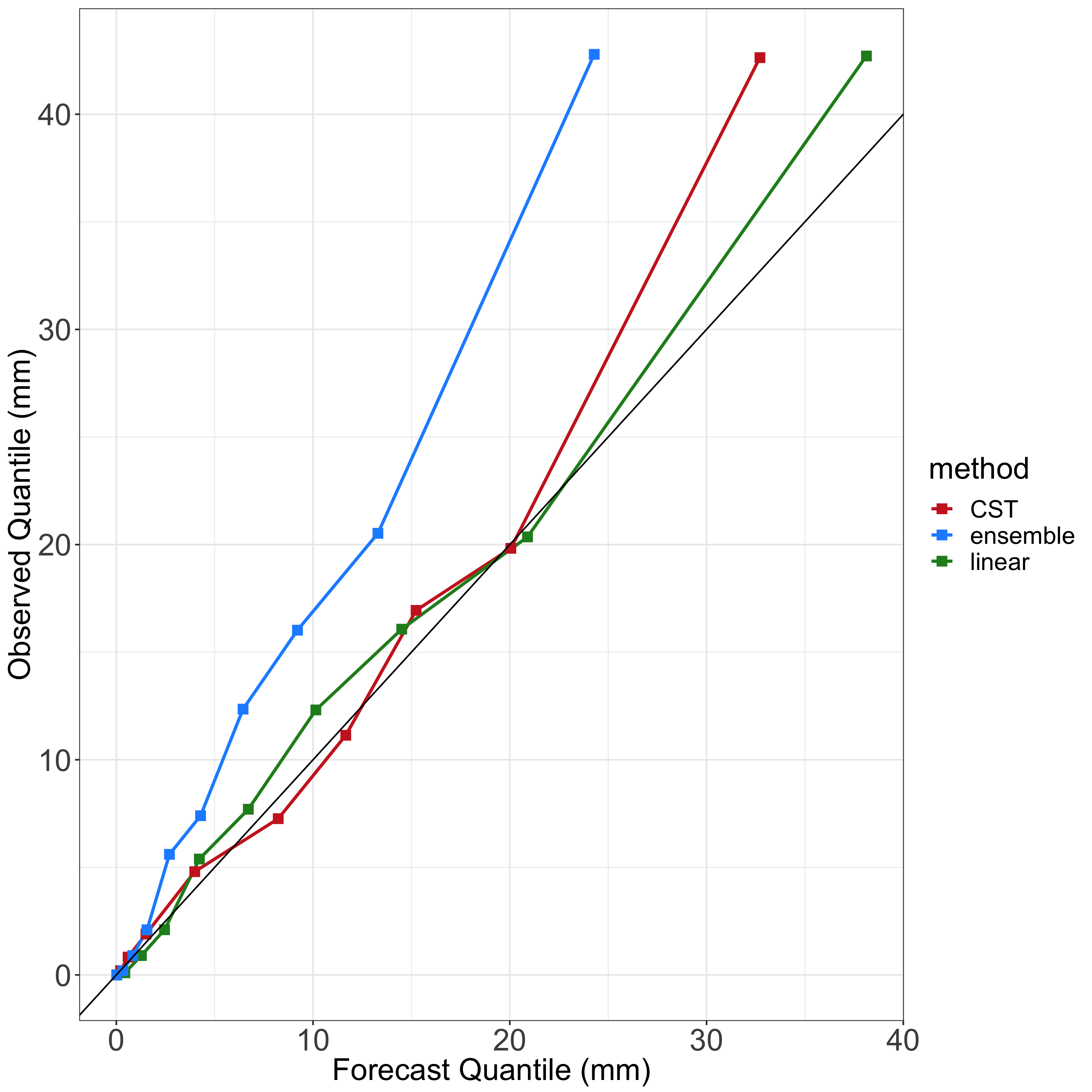}
	\includegraphics[width=0.45\textwidth]{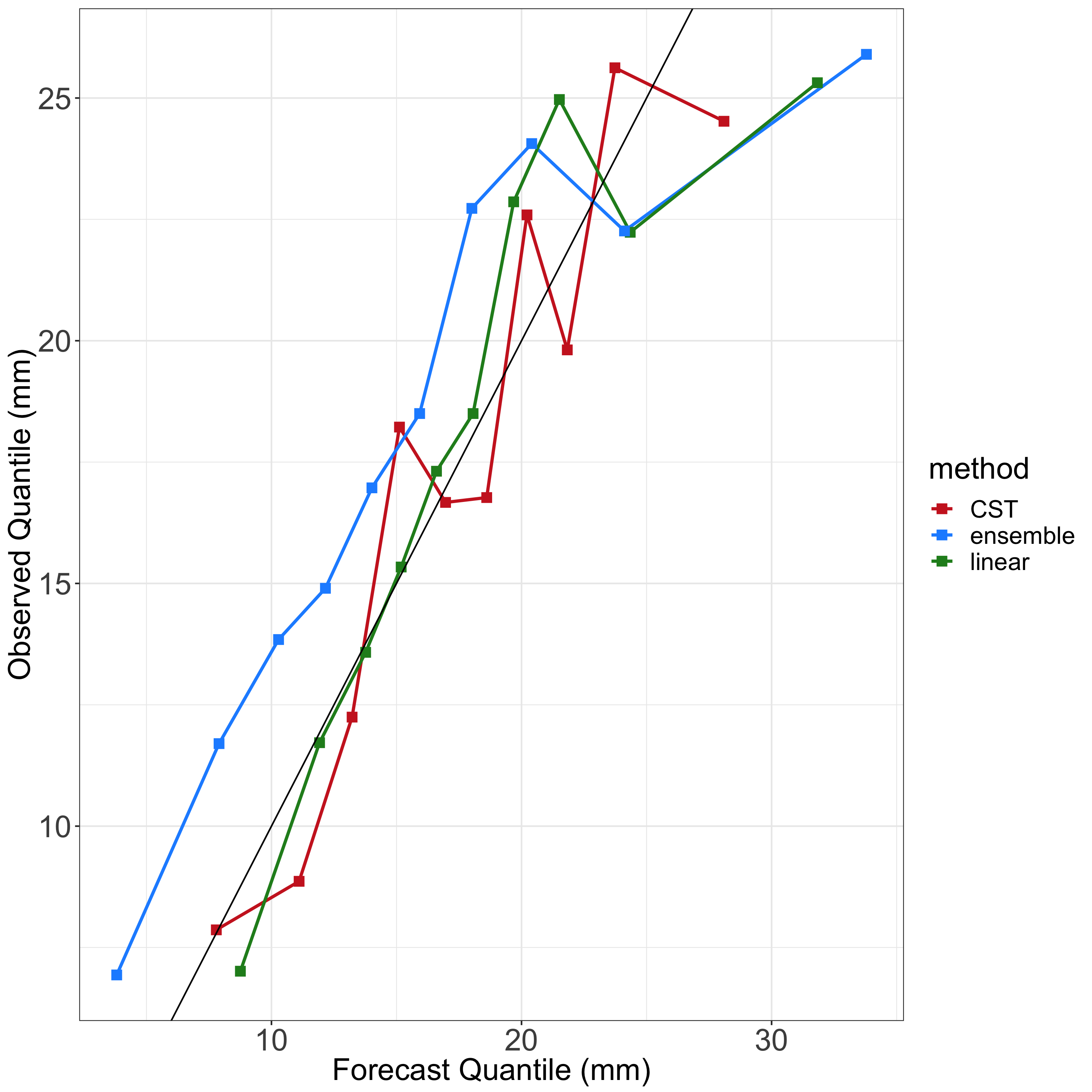}
	\caption{Quantile reliability diagrams for the CST estimator in red, the ensemble in blue and the linear estimator in green for the $\frac{51}{52}$ quantile; on the left side for 24 hours lead time and on the right side 192 hours lead time.}
	\label{fig:data-QRD}
\end{figure*}

The validation is carried out using a seven-fold cross validation, where, in every iteration, one year is left out of the model estimation and used as the independent validation sample. In Figure \ref{fig:data-QVSS} the QVSS is shown as a function of lead time. The bands are obtained by calculating the QVSS for each location separately. The graph on the left shows the performance of the CST estimator in red\jasper{, the linear estimator in green}  and the ensemble in blue for $\tau = \frac{51}{52}$. It can be observed that {\jasperv the CST and the linear estimator improve upon the ensemble especially for short lead times and for very long lead times. On the right side of the figure the performance of the $\tau=0.995$ quantile is shown for the CST and the linear estimators, showing that skilful quantile estimates are obtained up till 144 hours. The CST estimator seems to have slightly less spread in the scores than the linear method.

In practice the quantile estimates are of interest when the ensemble is already high, i.e. $X>t$ for $t$ large. In Figure \ref{fig:data-QVSS-thresh} similar plots are shown as in Figure \ref{fig:data-QVSS}, but now the verification is done based on a subset of the data where we condition on $X>5$, which is the 60 percent quantile for a lead time of 24 hours. Note that this means that also the reference climatological quantile has this conditioning. It can be seen in the left panel of Figure \ref{fig:data-QVSS-thresh} that the ensemble method is outperformed by the CST and the linear estimator for shorter lead times. For the extrapolation to $\tau=0.995$ in the right panel of Figure \ref{fig:data-QVSS-thresh}, the spread in skill of the different stations is much larger, but still showing skilful forecasts for most stations for short lead times. Also here the CST appears to have less spread than the linear estimator. In Figure \ref{fig:data-QRD} two quantile reliability diagrams are shown, for 24 hours lead time on the left and 192 hours lead time on the right, using all data without conditioning. The ensemble clearly underestimates the extremes generally for both lead times. The CST and the linear estimators improve calibration for 24 hour lead time. For a lead time of 192 hours the CST estimator looks a bit more unstable, though  it remains close the the calibration line, where the ensemble is consistently underestimating the upper quantile.

From all plots it can be concluded that the CST and the linear estimator are very comparable, an assumption of linear quantiles is in this context also not strange. Even though the CST estimator has a more flexible assumption on the quantile curves, it does not influence the results.}

\jasper{To conclude, we have shown that the CST estimator is comparable to the linear estimator and has more skill than the upper ensemble member for both short and long lead times. Additionally, it is able to extrapolate further into the tail and obtains skilful estimates for higher quantiles than are available from the ensemble.}

%
%

\section{Discussion} \label{sec:discussion}
{ \jasperv
We have estimated the conditional tail quantile curves, $x \mapsto Q_{Y|X}$, using a two step procedure. First we use local linear quantile regression to estimate a non-stationary threshold and secondly, extrapolate to the tail using the exceedances of this threshold. The assumption that $\gamma>0$ fits to the application of summer precipitation in the Netherlands, which is heavy tailed. There is a clear motivation for extending the model to the cases of light tailed, $\gamma = 0$  and short tailed data, $\gamma <0$. This would enable also post-processing of extreme precipitation in winter periods, but also temperature, wind speed and gusts and other weather phenomena. 

It is clear from the simulation that the linear method from \cite{Wang2012} is better able to deal with heteroskedastic data. Extending the model to allow for non-homoskedastic errors would be a valuable addition, allowing it to model data from a wider range of classes.

Finally, in the application we  now calibrate tail quantiles of the ensemble, using the statistical relation between the upper ensemble member and the observations. It would be of interest though, to consider a wider range of covariates from the NWP model. It would therefore be of value to extend the method to a multivariate covariates setting.}

\section*{Acknowledgements}
The authors would like to sincerely thank the two referees and the associate editor for the constructive comments which led to a substantial improvement of this paper.
This work is part of the research project ``Probabilistic forecasts of extreme weather utilizing advanced methods from extreme value theory" with project number 14612 which is financed by the Netherlands Organisation for Scientific Research (NWO).

\bibliographystyle{imsart-number}
\bibliography{bibliography}

\appendix

\section{Proofs}
This section contains the proofs of Theorems \ref{thm:consistency}-\ref{thm:quantile}  in Section \ref{sec:asymp}. Throughout this section, $c, c_1, c_2,\ldots$ denote positive constants, which are not necessarily the same at each occurrence.

\subsection{Proof of Theorem \ref{thm:consistency}}
The uniform consistency of $\hat r$ relies heavily on the uniform Bahadur representation for $\hat r$.  We make use of the Bahadur representation obtained in \cite{Kong2010}. 

Let  $\psi_{\tau}(u) = \tau-I(u<0)$, that is the right derivative of  $\rho_{\tau}$ at $u$. Then by Corollary 3.3 and Proposition 1  in \cite{Kong2010}, we have
			\begin{align*}
				&\sup_{x \in [a,b]} \left|		\hat{r}(x) - r(x) + 
				h_n^2 c r''(x) - 
				\frac{1}{nh_n}\sum_{i=1}^n \psi_{\tau_c}(\epsilon_i)C_{n, i}(x)K\left(\frac{X_i-x}{h_n}\right) \right|\\
				&= O_p\left(\left\{\frac{\log n}{nh_n}\right\}^{3/4}\right)
				= O_p\left(\left\{\frac{\log n}{n^{1-\delta_h}}\right\}^{3/4}\right),
			\end{align*}
where $C_{n, i}(x)$ is a Lipschitz continuous function and thus absolutely bounded in $[a,b]$.
Define 
\begin{equation*} 
 \Delta_n(x) =  \frac{1}{nh_n}\sum_{i=1}^n \psi_{\tau_c}(\epsilon_i)C_{n,i}(x)K\left(\frac{X_i-x}{h_n}\right).
 \end{equation*}
Then, the triangle inequality leads to 
\begin{align}\label{eq:r_unif}
\sup_{x \in [a,b]} \left|\hat{r}(x) - r(x) \right|\leq &\sup_{x \in [a,b]} \left|h_n^2 c r''(x)\right|+ \sup_{x \in [a,b]} \left|\Delta_n(x)\right|+ O_p\left(\left\{\frac{\log n}{n^{1-\delta_h}}\right\}^{3/4}\right) \nonumber\\
=&O(n^{-2\delta_h})+\sup_{x \in [a,b]} \left|\Delta_n(x)\right|+ O_p\left(\left\{\frac{\log n}{n^{1-\delta_h}}\right\}^{3/4}\right). 
\end{align}
The last equality follows from the fact that $r''$ is uniformly bounded by Assumption A1. 

Next, we show that, there exists a $\delta_C\in (0, \frac{1}{2}-\delta_h)$ such that 
\begin{equation}\label{eq:Delta_n}
\sup_{x \in [a,b]} \left|\Delta_n(x)\right| = o_p(n^{-\delta_C}).  
\end{equation}
Define $ T_i(x):= h_n K\left(\frac{X_i-x}{h_n}\right)C_{n, i}(x)$. Then for any $x, y \in [a, b]$, by the triangle inequality and the Lipschitz continuity of $K$, we have
\begin{align*}
	 &\left|T_i(x)-T_i(y)\right|= h_n \left|K\left(\frac{X_i-x}{h_n}\right)C_{n, i}(x) - K\left(\frac{X_i-y}{h_n}\right)C_{n,i}(y)\right| \\
	\leq &h_n |C_{n, i}(x)|\left|K\left(\frac{X_i-x}{h_n}\right) - K\left(\frac{X_i-y}{h_n}\right)\right| + h_nK\left(\frac{X_i-y}{h_n}\right)|C_{n, i}(x) - C_{n,i}(y)|\\
	\leq& c_1\left|x-y\right| + c_2 h_n |x-y|\sup_{u\in [-1,1]}K(u)\\
	\leq& c|x-y|.
			\end{align*}
Note that the constant $c$ does not depend on $i$, that is, the Lipschitz continuity is uniform in $i$ for all $T_i$'s. Consequently, it follows from that $\left|\psi_{\tau}(u)\right| \leq 1$ that,
\begin{align*}
	|\Delta_n(x) - \Delta_n(y)| = \frac{1}{nh^2_n}\left|\sum_{i=1}^n \psi_{\tau_c}(\epsilon_i)(T_i(x) - T_i(y)) \right|
				\leq c\frac{|x-y|}{h_n^2}.
			\end{align*}
Let $M_n = n^{\delta_C+2\delta_h}\log n$ and $\left\{ I_i=(t_i,t_{i+1}], i=1,\ldots , M_n \right\}$ be a partition of $(a,b]$, where $t_{i+1} - t_{i} = \frac{b-a}{M_n}$. Then for $t \in I_i$,
			\begin{align*}
				|\Delta_n(t) - \Delta_n(t_i)| \leq \frac{c(b-a)}{M_n h_n^2},
			\end{align*}
or equivalently,
			\begin{align*}
				\Delta_n(t_i) - \frac{c(b-a)}{M_n h_n^2} \leq \Delta_n(t) \leq \Delta_n(t_i) + \frac{c(b-a)}{M_n h_n^2}.
			\end{align*}
Therefore, for $n$ sufficiently large, 
			\begin{align*}
				&\PP\left(\sup_{x\in[a,b]} |\Delta_n(x)| > n^{-\delta_C} \right) &=& \PP\left(\max_{1\leq i \leq M_n} \sup_{t\in I_i}|\Delta_n(t)| > n^{-\delta_C} \right)\\
				&\leq \sum_{i=1}^{M_n} \PP\left(\sup_{t\in I_i} |\Delta_n(t)| > n^{-\delta_C} \right)
				&\leq& \sum_{i=1}^{M_n} \PP\left(|\Delta_n(t_i)| > n^{-\delta_C} - \frac{c(b-a)}{M_n h_n^2}\right)\\
				&\leq\sum_{i=1}^{M_n} \PP\left(|\Delta_n(t_i)| > \frac{1}{2}n^{-\delta_C}  \right) 
	&=&\sum_{i=1}^{M_n} \PP\left(\left| \sum_{j=1}^n \frac{T_j(t_i)\psi_{\tau_c}(\epsilon_j)}{h_n} \right| > \frac{1}{2}h_nn^{1-\delta_C} \right) =: \sum_{i=1}^{M_n} P_i,
			\end{align*}
where the third inequality is due to that $\frac{c(b-a)}{M_n h_n^2}<\frac{1}{2}n^{-\delta_C}$ for $n$ sufficiently large. Next, we apply Hoeffding's inequality to bound $P_i$. Define $$W_{n,i,j}:= \frac{T_j(t_i)\psi_{\tau_c}(\epsilon_j)}{h_n}= K\left(\frac{X_j-t_i}{h_n}\right)C_{n, j}(t_i) \psi_{\tau_c}(\epsilon_j).$$
 For each $i$ and $n$, $\{W_{n,i,j}, 1\leq j\leq n\}$ is a sequence of i.i.d. random variables. And with probability one, $|W_{n,j,i}|\leq \sup_{-1\leq u\leq1}K(u)\sup_{a\leq x\leq b}C_{n,i}(x)=:c_3$. Moreover, $\EE\left( W_{n,j,i} \right)=0$ because  $\EE(\psi_{\tau_c}(\epsilon_j))=0$ and $X_j$  and $\epsilon_j$ are independent. Thus,  by Hoeffding's inequality,
 \begin{align*}
P_i= \PP\left(\left| \sum_{j=1}^n W_{n,i,j}\right| \geq \frac{1}{2}h_nn^{1-\delta_C}\right)\leq 2\exp\left(-\frac{n^{1-2\delta_C} h_n^2}{8c_3^2} \right)=2\exp\left(-cn^{ 1- 2 \delta_h -2\delta_C}\right).
\end{align*}
Note that $1- 2 \delta_h -2\delta_C>0$ by the choice of $\delta_C$. Thus,
 for $n \to \infty$, 		 
\begin{equation*}
\PP\left(\sup_{x\in[a,b]} |\Delta_n(x)| > n^{-\delta_C}\right) \leq 2M_n\exp\left(-cn^{1-2\delta_h-2\delta_C}\right)\rightarrow 0.
\end{equation*}
Hence, \eqref{eq:Delta_n} is proved. Now by choosing $\delta = \delta_C$, we obtain via \eqref{eq:r_unif} that,
\begin{equation*}
	n^{\delta} \sup_{x\in [a,b]} | \hat{r}_n(x) - r(x)| 
	= O(n^{\delta_C-2\delta_h})
	+ o_p(1)
	+ O_p\left(n^{-\frac{3}{4} + \frac{3}{4}\delta_h +\delta_C}(\log n)^{\frac{3}{4}}\right)
	= o_p(1),
\end{equation*} 
due to that $\delta_h\in(\frac{1}{5}, \frac{1}{2})$ and $\delta_C<\frac{1}{2}-\delta_h$.

\subsection{Proof of Theorem \ref{thm:gamma}}


The proof follows a similar line of reasoning as that of Theorem 2.1 in \cite{Wang2012}. The uniform consistency of $\hat r_n$ given in Theorem \ref{thm:consistency} plays a crucial role. Define $V_n \coloneqq ||\hat{r}_n-r||_{\infty} = o_p\left(n^{-\delta}\right)$.

Let $U_i=F_{Y|X}(Y_i|X_i)$ for all $1\leq i \leq n$. Then $\{U_i,  i=1,\ldots,n\}$ constitute i.i.d. random variables from a standard uniform distribution. 
{\juan{Recall the definition of $e_i$: 
$$
e_i=Y_i-\hat r_n(X_{i})=Q_{Y|X}(U_i|X_i)-\hat r_n(X_{i}).
$$
Thus, the ordering of $\{e_i, i=1,\ldots,n\}$ is not necessarily the same as the ordering of $\{U_i, i=1,\ldots,n\}$. The main task of this proof is to show that the $k_n$ largest $e_i$'s correspond to the  $k_n$ largest $U_i$'s; see \eqref{eq: es}.}} To this aim, we first prove that with probability tending to one, $e_{n-j,n}$ for $j=0,\ldots , k_n$ can be decomposed as follows,
\begin{equation} \label{eq:upp-residuals}
	e_{n-j,n} = Q_{\epsilon}(U_{i(j)}) + r(X_{i(j)})-\hat r_n(X_{i(j)}) \mbox{  for } j=0,\ldots k_n ,
\end{equation}
where $i(j)$ is the index function defined as $e_{i(j)} = e_{n-j,n}$. In view of  \eqref{eq:residuals}, it is sufficient to prove that with probability tending to one, $U_{i(j)} > \tau_c$  jointly for all $j=0,\ldots ,k_n$.  Define another index function, $\tilde{i}(j)$ by $U_{\tilde{i}(j)} = U_{n-j,n}$. Then it follows for $n$ large enough,
\begin{align*}
	\PP \left(\cup_{j=0}^{k_n} \{ U_{i(j)} < \tau_c \}\right) 
	&= \PP\left(\cup_{j=0}^{k_n} \{ Y_{i(j)} < Q_{Y|X}(\tau_c|X_{i(j)})\} \right)\\
	&= \PP\left(\min_{0\leq j \leq k_n} \left(Y_{i(j)} - r(X_{i(j)})\right) < 0 \right)\\
	&= \PP\left(\min_{0\leq j \leq k_n}\left( Y_{i(j)} - \hat{r}_n(X_{i(j)}) - r(X_{i(j)}) + \hat{r}_n(X_{i(j)})\right) < 0 \right)\\
	&\leq \PP\left(\min_{0\leq j \leq k_n} e_{n-j,n} - \sup_{x \in [a,b]}|\hat{r}_n(x) - r(x)| < 0 \right)\\
	&= \PP\left(e_{n-k_n,n} < V_n \right) 
	 = 1-\PP(e_{n-k_n,n}\geq V_n)\\
	&\leq 1- \PP\left(\cap_{j=0}^{k_n} \{ e_{\tilde{i}(j)} \geq V_n \} \right)\\
	&= 1- \PP\left(\cap_{j=0}^{k_n} \left\{ Q_{\epsilon}(U_{n-j,n}) + r(X_{\tilde{i}(j)}) - \hat{r}_n(X_{\tilde{i}(j)}) \geq V_n \right\} \right)\\
	&\leq 1- \PP\left( Q_{\epsilon}(U_{n-k_n,n}) \geq 2V_n \right),
\end{align*}
where the second equality follows from that $Q_{Y|X}(\tau_c|X_{i(j)})=r(X_{i(j)})$ and the last equality follows from \eqref{eq:residuals} and the fact that
 $U_{n-k_n,n} > \tau_c$ for $n$ large enough. Then, 
$\lim_{n\to\infty} \PP \left(\cup_{j=0}^{k_n} \{U_{i(j)} < \tau_c \}\right) = 0$ follows from $Q_{\epsilon}(U_{n-k_n,n}) \to \infty$ and $V_n = o_p(1)$ as $n\to \infty$. Hence, \eqref{eq:upp-residuals} is proved.
	
Next, we show that 
\begin{equation}
\lim_{n\rightarrow\infty}\PP\left(\cap_{j=0}^{k_n} \{e_{n-j,n}= Q_{\epsilon}(U_{n-j,n})+ r(X_{i(j)})-\hat r_n(X_{i(j)})\}\right)=1,   \label{eq: es}
\end{equation}
 that is the ordering of $k$ largest residuals is determined by the ordering of $U_i$'s. In view of \eqref{eq:upp-residuals}, it is sufficient to show that with probability tending to one,
\begin{equation} \label{eq:Q_e}
\min_{1 \leq i \leq k_n} (Q_{\epsilon}(U_{n-i+1,n})-Q_{\epsilon}(U_{n-i,n}))\geq 2\max_{1 \leq i \leq k_n}|r(X_{i(j)})-\hat r_n(X_{i(j)}|.
\end{equation}
By the second order condition given in \eqref{eq:sec-ord-cond} and Theorem 2.3.9 in \cite{Haan2007}, for any small $\delta_1, \delta_2>0$, and $n$ large enough,
{\juan{\begin{align}
	\frac{Q_{\epsilon}(U_{n-i+1,n}) }{Q_{\epsilon}(U_{n-i,n})}
	\geq W_i^\gamma+A_0\left(\frac{1}{1-U_{n-i,n}}\right)W_i^\gamma\frac{W_i^\rho-1}{\rho}-\delta_1\left|A_0\left(\frac{1}{1-U_{n-i,n}}\right)\right|W_i^{\gamma+\rho+\delta_2}, \label{eq: lo_bound}
\end{align}}}
for $i=1,\ldots,k_n$, where $W_i=\frac{1-U_{n-i,n}}{1-U_{n-i+1,n}}$ and $\lim_{t\rightarrow\infty}A_0(t)/A(t)=1$. Observe that $\log W_i= \log \frac{1}{1-U_{n-i+1,n}}-\log \frac{1}{1-U_{n-i,n}}  \overset{d}{=}E_{n-i+1,n}-E_{n-i,n}$ with $E_i$'s i.i.d. standard exponential variables.
 Thus,  by R\`enyi's representation \cite{Renyi1953},  we have
\begin{equation*}
\{W_i, 1\leq i\leq k_n\}\overset{d}{=}\left\{\exp\left(\frac{E_i}{i}\right), 1\leq i\leq k_n \right\}.
\end{equation*}

{\juan{From Proposition 2.4.9 in \cite{Haan2007},  we have $\frac{U_{n-k_n,n}}{1-\frac{k_n}{n}}\overset{P}{\rightarrow}1$, which implies that ${A_0\left(\frac{1}{1-U_{n-k_n,n}}\right)}=O_p\left({A_0\left(\frac{n}{k_n}\right)}\right)$. }}Using the fact that $A_0$ is regularly varying with index $\mathcal{\rho}$, hence  $|A_0|$ is ultimately decreasing, we obtain for $n$ sufficiently large and any $i = 1,\ldots, k_n$,
{\juan{\begin{align}
\left|A_0\left(\frac{1}{1-U_{n-i,n}}\right)\right|\leq& \left|A_0\left(\frac{1}{1-U_{n-k_n,n}}\right)\right| \nonumber\\
=&\left|O_p\left(A_0\left(\frac{n}{k_n}\right)\right)\right|=\left|O_p\left(A\left(\frac{n}{k_n}\right)\right)\right|=\left|O_p\left(\frac{1}{\sqrt{k_n}}\right)\right|,  \label{eq: bound_A}
\end{align}
}}
by the assumption $\sqrt{k_n}A\left(\frac{n}{k_n}\right)\rightarrow \lambda$. 

For a sufficiently large $u$ and any $k_n\geq 1$,
\begin{align*}
&\PP\left(\max_{1\leq i\leq k_n}\frac{E_i}{i}\leq u\right)=\prod_{i=1}^{k_n} \left(1-e^{-iu}\right)=\exp\left(\sum_{i=1}^{k_n}\log  \left(1-e^{-iu}\right) \right)\\
=&\exp\left(-\sum_{i=1}^{k_n}\sum_{j=1}^\infty j^{-1}e^{-iuj} \right)\geq \exp\left(-\sum_{i=1}^{k_n}e^{-iu}\right)=
\exp\left( \frac{1-e^{-ku}}{1-e^u}\right),
\end{align*}
which tends to one as $u\rightarrow \infty$. This implies that 
\begin{align}
\min_{1\leq i\leq k_n}W_i^\rho\overset{d}{=}\exp\left(\rho \max_{1\leq i\leq k_n}\frac{E_i}{i}\right)=O_p(1). \label{eq: bound_Wi}
\end{align}

 Thus, combining \eqref{eq: lo_bound}, \eqref{eq: bound_A} and \eqref{eq: bound_Wi}, we have
{\juan{
\begin{align*}
&\min_{1\leq i\leq k_n}\frac{Q_{\epsilon}(U_{n-i+1,n}) }{Q_{\epsilon}(U_{n-i,n})}-1 \\
\geq &\min_{1\leq i\leq k_n}W_i^\gamma\left(1-\left|O_p\left(\frac{1}{\sqrt{k_n}}\right) \right| \left( \frac{W_i^\rho-1}{\rho}+  \delta_1W_i^{\rho+\delta_2} \right)\right)-1\\
= &\min_{1\leq i\leq k_n}W_i^\gamma\left(1-\left|O_p\left(\frac{1}{\sqrt{k_n}}\right)\right|\right)-1 
\overset{d}{=}\exp\left(\gamma\frac{E_1}{k_n}\right)\left(1-\left|O_p\left(\frac{1}{\sqrt{k_n}}\right)\right|\right)-1\\
=&\frac{\gamma E_1}{k_n}\left(1-\left|O_p\left(\frac{1}{\sqrt{k_n}}\right)\right|\right),
\end{align*}
}}
where  the third equality follows from that $\min_{1\leq i\leq k_n}\frac{E_i}{i}\overset{d}{=}E_{1,k}\overset{d}{=}\frac{E_1}{k}$ by R\`enyi's representation.
Thus, we obtain that 
\begin{align*}
	\min_{1 \leq i \leq k_n} (Q_{\epsilon}(U_{n-i+1,n})-Q_{\epsilon}(U_{n-i,n}))
	&\geq \left(Q_{\epsilon}(U_{n-k_n,n})\frac{\gamma E_1}{k_n}\right)\left(1-\left|O_p\left(\frac{1}{\sqrt{k_n}}\right)\right|\right) \\
&= \left(\frac{n}{k_n}\right)^\gamma k_n^{-1}|O_p(1)|.
\end{align*}
Thus, \eqref{eq:Q_e} is proved by the assumption $k_n^{-1}\left(\frac{n}{k_n}\right)^{\gamma}>>n^{-\delta}$ and $\max_{1 \leq i \leq k_n}|r(X_{i(j)})-\hat r_n(X_{i(j)}|\leq 2 V_n=o_p\left(n^{-\delta}\right)$.
Intuitively, \eqref{eq:Q_e} means that the difference between two successive upper order statistics of $\epsilon$ is larger than the error made in the estimation of $r(x)$. 

As aforementioned,  \eqref{eq:upp-residuals} and \eqref{eq:Q_e} together lead to  \eqref{eq: es}, which further implies that with probability tending to one,

	\begin{equation} \label{eq:residual-expression}
		\max_{0 \leq j \leq k_n} \left|\frac{e_{n-j,n}}{Q_{\epsilon}(U_{n-j,n})}-1 \right| \leq \frac{V_n}{Q_{\epsilon}(U_{n-k_n,n})}
=o_p\left(n^{-\delta}\left(\frac{k_n}{n}\right)^\gamma\right).
	\end{equation}

By  the definition of $\hat{\gamma}_n$ and \eqref{eq:residual-expression}, we can write the estimator as follows,
\begin{align*}
	\hat{\gamma}_n &= \frac{1}{k_n} \sum_{i=0}^{k_n-1}\log \frac{e_{n-i,n}}{e_{n-k_n,n}}\\
&= \frac{1}{k_n} \sum_{i=0}^{k_n-1} \log\frac{Q_{\epsilon}(U_{n-i,n})}{Q_{\epsilon}(U_{n-k_n,n})} 
	+ \left(\frac{1}{k_n} \sum_{i=0}^{k_n-1}\log \frac{e_{n-i,n}}{Q_{\epsilon}(U_{n-i,n})}-\log \frac{e_{n-k_n,n}}{Q_{\epsilon}(U_{n-k_n,n})}\right)\\
	&=: \hat{\gamma}_H + o_p\left(n^{-\delta}\left(\frac{k_n}{n}\right)^\gamma\right).
	\end{align*}
	The first part is the well known Hill estimator and we have by Theorem 3.2.5 in \cite{Haan2007},
	\begin{align*}
		\sqrt{k_n}(\hat{\gamma}_H-\gamma)) \xrightarrow{d} N\left( \frac{\lambda}{1-\varrho},\gamma^2\right).
\end{align*}
Therefore we can conclude,
\begin{align*}
			\sqrt{k_n}(\hat{\gamma}_n - \gamma) = \sqrt{k_n}(\hat{\gamma}_H-\gamma) +o_p\left(\sqrt{k_n}n^{-\delta}\left(\frac{k_n}{n}\right)^\gamma\right)  \xrightarrow{d} N\left( \frac{\lambda}{1-\varrho},\gamma^2\right),
		\end{align*}
by the assumption that $k_n^{\gamma+1}n^{-\gamma-\delta}\rightarrow 0$.

We remark that the proof for Theorem 2.1 in \cite{Wang2012} isn't completely rigorous, namely, the proof for (S.1) in the supplementary material of that paper is not right. We fix the problem while proving \eqref{eq:residual-expression}, which is an analogue to (S.1).

\subsection{Proof of Theorem \ref{thm:epsilon}}
Before we proceed with the proof of Theorem \ref{thm:quantile}, we state the asymptotic normality of $\hat{Q}_{\epsilon}(\tau_n)$ defined in \eqref{eq:epsilon-estimator} in the theorem below.
	\begin{thm}\label{thm:epsilon}
	 Let the conditions of Theorem \ref{thm:gamma} be satisfied. Assume $np_n = o(k_n)$ and $\log(np_n) = o(\sqrt{k_n})$, then, as $n \to \infty$,
	\begin{equation} \label{eq:quant-norm}
		\frac{\sqrt{k_n}}{\log(k_n/(np_n))} \left( \frac{\hat{Q}_{\epsilon}(\tau_n)}{Q_{\epsilon}(\tau_n)} -1 \right) \xrightarrow{d} N\left( \frac{\lambda}{1-\varrho},\gamma^2\right).
	\end{equation}
	\end{thm}

Theorem \ref{thm:epsilon} can be proved in the same way as that for Theorem 2 in \cite{Wang2012}. For the sake of completeness, we present the proof in this section.

Recall that $\hat{Q}_{\epsilon}(\tau_n) = \left( \frac{k_n}{np_n} \right)^{\hat{\gamma}_n} e_{n-k_n,n}=:d_n^{\hat{\gamma}_n} e_{n-k_n,n}$. First, note that from Theorem \ref{thm:gamma}, we have $\sqrt{k_n}(\hat{\gamma}_n-\gamma)=\Gamma+o_p(1)$, where $\Gamma$ is a random variable from $N\left( \frac{\lambda}{1-\varrho},\gamma^2\right)$. Therefore,
\begin{align}
d_n^{\hat{\gamma}_n-\gamma}=&\exp\left((\hat{\gamma}_n-\gamma)\log d_n\right)=\exp\left(   \frac{\log d_n}{\sqrt{k_n}} (\Gamma+o_p(1))  \right) \nonumber\\
=&1+ \frac{\log d_n}{\sqrt{k_n}}\Gamma+o_p(\frac{\log d_n}{\sqrt{k_n}}), \label{eq: hat_gamma}
\end{align}
where the last step follows from the assumption that $\frac{\log d_n}{\sqrt{k_n}} \to 0$. Second, by Theorem 2.4.1, 
$$
\sqrt{k}\left( \frac{Q_{\epsilon}(U_{n-k_n,n}) }{Q_{\epsilon}(1-k_n/n)}-1\right)\xrightarrow{d} N(0,\gamma^2).
$$
In combination with \eqref{eq:residual-expression}, we have 
\begin{align}
\frac{e_{n-k_n,n}}{Q_{\epsilon}(1-k_n/n)}=&\frac{e_{n-k_n,n}}{Q_{\epsilon}(U_{n-k_n,n}) }\cdot\frac{Q_{\epsilon}(U_{n-k_n,n})}{Q_{\epsilon}(1-k_n/n)}  
=\left(1+o_p\left(n^{-\delta}\left(\frac{k_n}{n}\right)^\gamma\right)\right)\left(1+O_p\left(\frac{1}{\sqrt{k_n}}\right)\right) \nonumber\\
=&1+O_p\left(\frac{1}{\sqrt{k_n}}\right),  \label{eq: eQ}
\end{align}
by the assumption that $k_n^{\gamma+1}n^{-\gamma-\delta}\rightarrow 0$. Last, by the second order condition given in \eqref{eq:sec-ord-cond} and Theorem 2.3.9 in \cite{Haan2007},
\begin{align}
\frac{Q_{\epsilon}(1-p_n)}{Q_{\epsilon}(1-k_n/n) d_n^{\gamma}}=1+O(A(n/k_n))=1+O\left(\frac{1}{\sqrt{k_n}}\right). \label{eq: Qp}
\end{align}

Finally, combing \eqref{eq: hat_gamma}, \eqref{eq: eQ} and \eqref{eq: Qp}, we have 
\begin{align*}
		\frac{\hat{Q}_{\epsilon}(\tau_n)}{Q_{\epsilon}(\tau_n)} =&\frac{d_n^{\hat{\gamma}}e_{n-k_n,n}}{Q_{\epsilon}(1-p_n)}
=d_n^{\hat{\gamma}_n-\gamma}\frac{e_{n-k_n,n}}{Q_{\epsilon}(1-k_n/n)}\cdot \frac{Q_{\epsilon}(1-k_n/n) d_n^{\gamma}}{Q_{\epsilon}(1-p_n)}\\
=&\left(1+ \frac{\log d_n}{\sqrt{k_n}}\Gamma+o_p\left(\frac{\log d_n}{\sqrt{k_n}}\right)\right)\left(  1+O_p\left(\frac{1}{\sqrt{k_n}}\right)  \right)\left(  1+O\left(\frac{1}{\sqrt{k_n}}\right)\right)\\
=&1+ \frac{\log d_n}{\sqrt{k_n}}\Gamma+o_p\left(\frac{\log d_n}{\sqrt{k_n}} \right),
\end{align*}
by the assumption that $d_n\rightarrow\infty$. Thus, \eqref{eq:quant-norm} follows immediately.

\subsection{Proof of Theorem \ref{thm:quantile}}
By definition of $\hat{Q}_{Y|X}(\tau_n|x)$ and Theorem \ref{thm:consistency}, we have,
\begin{align*}
&\frac{\sqrt{k_n}}{\log\left(\frac{k_n}{np_n}\right)Q_{\epsilon}(\tau_n)}\left(\hat{Q}_{Y|X}(\tau_n|x) - Q_{Y|X}(\tau_n|x)\right)\\ 
=&\frac{\sqrt{k_n}}{\log\left(\frac{k_n}{np_n}\right)Q_{\epsilon}(\tau_n)}
\left( \hat{Q}_{\epsilon}(\tau_n) - {Q}_{\epsilon}(\tau_n)+\hat r_n(x)-r(x) \right),\\
=&\frac{\sqrt{k_n}}{\log\left(\frac{k_n}{np_n}\right)Q_{\epsilon}(\tau_n)}
\left( \hat{Q}_{\epsilon}(\tau_n) - {Q}_{\epsilon}(\tau_n)\right)+O_p\left(\frac{\sqrt{k_n}n^{-\delta}}{\log\left(\frac{k_n}{np_n}p_n^{-\gamma}\right)}\right).
\end{align*}
Thus it follows from Theorem \ref{thm:epsilon} and the assumption $\frac{\sqrt{k}p_n^{\gamma}}{n^{\delta}\log \left(\frac{k_n}{np_n}\right)} \to 0$ that 
				\begin{align*}
					\frac{\sqrt{k_n}}{\log\left(\frac{k_n}{np_n}\right)Q_{\epsilon}(\tau_n)}\left(\hat{Q}_{Y|X}(\tau_n|x) - Q_{Y|X}(\tau_n|x)\right) \xrightarrow{d} N\left(\frac{\lambda}{1-\varrho},\gamma^2\right).
				\end{align*}
\end{document}